\documentclass[11pt,towside,floatfix]{article}
\usepackage{amsmath,amssymb,epsf,graphicx,cite,bbm}
\begin{document}

\newcommand{\mathbold}[1]{\mbox{{\boldmath $#1$}}}
\newcommand{\beq}{\begin{equation}}
\newcommand{\eeq}{\end{equation}}
\newcommand{\beqn}{\begin{eqnarray}}
\newcommand{\eeqn}{\end{eqnarray}}
\newcommand{\bearr}{\begin{array}}
\newcommand{\enarr}{\end{array}}
\newcommand{\derp}[2]{\frac{\partial{#1}}{\partial{#2}}}
\newcommand{\toref}[1]{\mbox{(\ref{#1})}}
\newcommand{\ket}[1]{|#1\rangle}
\newcommand{\bra}[1]{\langle#1|}
\newcommand{\bracket}[2]{\langle#1|#2\rangle}
\newcommand{\comm}[2]{\left[#1,#2\right]}

\newcommand{\anticomm}[2]{\left\{#1,#2\right\}}
\newcommand{\dcomm}[3]{\left[#1,#2,#3\right]}
\newcommand{\danticomm}[3]{\left\{#1,#2,#3\right\}}
\newcommand{\expval}[3]{\langle#1|#2|#3\rangle}
\newcommand{\redexpval}[3]{\langle#1||#2||#3\rangle}
\newcommand{\media}[1]{\langle#1\rangle}
\newcommand{\eps}{\varepsilon}

\title{Mean field theory for skewed height profiles\\ in KPZ growth processes}
\maketitle

\begin{center}
Francesco Ginelli and Haye Hinrichsen 

\vglue 3mm
Institut f{\"u}r Theoretische Physik und Astrophysik, \\
        Universit{\"a}t W{\"u}rzburg, D-97074 W{\"u}rzburg, Germany

\end{center}

\begin{abstract}
We propose a mean field theory for interfaces growing according to the 
Kardar-Parisi-Zhang (KPZ) equation in 1+1 dimensions. The mean field equations
are formulated in terms of densities at different heights, taking surface
tension and the influence of the nonlinear term in the KPZ equation into account.
Although spatial correlations are neglected, the mean field equations still reflect
the spatial dimensionality of the system. In the special case of Edwards-Wilkinson growth,
our mean field theory correctly reproduces all features. In the presence of a
nonlinear term one observes a crossover to a KPZ-like behavior with the correct
dynamical exponent $z=3/2$. In particular we compute the skewed interface profile
during roughening, and we study the influence of a co-moving
reflecting wall, which has been discussed recently in the context of nonequilibrium
wetting and synchronization transitions. Also here the mean field approximation 
reproduces all qualitative features of the full KPZ equation, although with different
values of the surface exponents.
\end{abstract}


\def\xvec{\vec{x}}                      
\def\nupar{{\nu_\parallel}}             
\def\nuperp{{\nu_\perp}}                

\section{Introduction}

Since many years the physical properties of deposition-evaporation processes on a planar
surface have been studied theoretically by analyzing appropriate 
stochastic growth models that capture the essential features of the
experimental realm~\cite{Dietrich}. In most of these models the configuration of 
the growing surface is described by a height variable $h(\xvec,t)$ that yields the height
of the interface between deposited layer and gas phase above point $\xvec$ of
the substrate at time $t$. Starting with a certain initial configuration, the 
interface then evolves according to certain stochastic rules.

Depending on the specific dynamic rules for deposition and evaporation and their
symmetries, the temporal evolution of the interface may be described on a
coarse-grained scale by a stochastic differential equation, one of the
simplest and more general one being the celebrated Kardar-Parisi-Zhang (KPZ) 
equation~\cite{KPZ}
\begin{equation}
\label{KPZeq}
\frac{\partial}{\partial t}h(\xvec,t) \;=\; v_0 + D \nabla^2 h(\xvec,t) 
+\frac{\lambda}{2} [\nabla h(\xvec,t)]^2 + \xi(\xvec,t)\,.
\end{equation}
Here $v_0$ is the average growth velocity which can be set to zero in a
co-moving frame, the Laplacian accounts for surface tension of the interface, 
and $\xi(\xvec,t)$ is an uncorrelated white Gaussian noise generated by the stochastic
nature of deposition and evaporation. Moreover, the nonlinear term $(\nabla h)^2$ 
it's the simplest one which breaks the invariance under reflections $h \to -h$. 

As many models for interface growth, the KPZ equation exhibits dynamic scaling,
i.e., starting with a flat configuration the interface width 
$w(t)=\sqrt{\langle h^2 \rangle - \langle h \rangle^2}$ (where $\langle \cdot \rangle$
denotes average over space and ensemble realizations) first increases as a
power law $w(t)\sim t^\gamma$ until it saturates in a finite system of linear 
size $L$ at a stationary value $w_{\text{stat}} \sim L^\alpha$. The crossover from a roughening to a 
stationary state is described by the well-known Family-Vicsek scaling form~\cite{FVScaling}
\begin{equation}
\label{VWScaling}
w(t) \;\sim\; t^\gamma \,g(t/L^z)\,,
\end{equation}
where $g$ is a universal scaling function and $z=\alpha/\gamma$ is the dynamical exponent. 

The width is actually related to the second moment of the height distribution
profile $P_L(h,t)$, which is defined as the normalized probability to find the
interface at a randomly chosen lattice site at height $h$. Clearly,
the height distribution contains much more information about the interface
morphology than the width alone. 
As for the width, dynamic scaling implies a scaling form for the height
distribution which in a co-moving frame may be written 
as\footnote{The scaling functions $g$ and $f$ are related by 
$g^2(u)=\int_0^\infty f(u,v)v^2\,dv-[\int_0^\infty f(v,u)v\,dv]^2$.}
\begin{equation}
\label{ScalingForm}
P_L(h,t) \;=\; t^{-\gamma} \, f(h/t^\gamma,t/L^z) \,.
\end{equation}
Obviously, both the critical exponents $\alpha$ and $\gamma$ and the shape of height
profile during roughening or after saturation reflect the symmetries of the
growth process under consideration. The simple case of invariance under
the reflection $h \to -h$ can be studied by imposing $\lambda = 0$ in
Eq. \toref{KPZeq}. In this case the linear Edwards-Wilkinson (EW) equation
\cite{EW} is recovered, and the critical growth exponents in 1+1 dimensions take the 
values $\alpha=1/2$ and $\gamma=1/4$. Moreover, the height profile 
of 1+1-dimensional EW processes is known to be a simple Gaussian distribution
both in the dynamically roughening phase as well as in the stationary state.

In more realistic growth models, where nearest neighbour interactions play a
role in the dynamics of the growing interface, reflection symmetry is broken
and the nonlinear KPZ term has to be taken into account. 
In 1+1 dimensions such a term is known to be {\it relevant} in the renormalization group
sense. Therefore, even when the
reflection symmetry is weakly violated (i.e., if $\lambda$ is small), 
the scaling behavior of an infinite system will eventually cross over from EW to
KPZ scaling, the latter being characterized by the 
exponents $\alpha=1/2$, $\gamma=1/3$, and $z=\alpha/\gamma=3/2$. 

\begin{figure}
\centerline{\includegraphics[width=105mm]{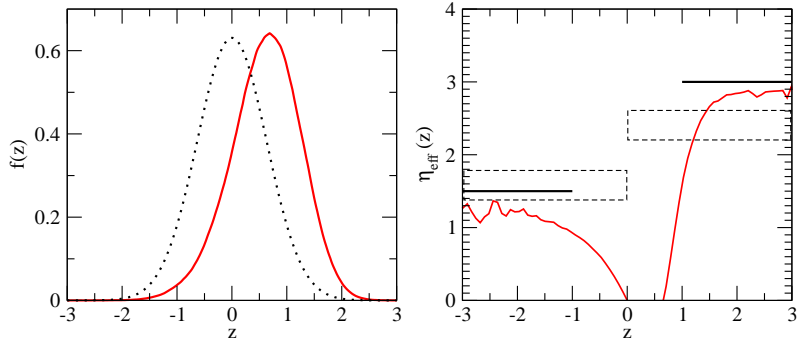}}
\caption{\label{figprofile} \small
Skewed interface profile. Left panel: Numerically determined profile $f(z):=
f(z,0)$ of a roughening KPZ interface in 1+1 dimensions for $\lambda<0$ (solid
line) compared to a Gaussian distribution of the same width (dashed
line). Right panel: Effective exponents $\eta_{eff}(z)$ (see text)
in comparison with the numerical estimates $\eta_+=1.6(2)$ and
$\eta_-=2.4(2)$ reported in Ref.~\cite{KBM91a} (representing the error bars as
dashed boxes) and the theoretical predictions for the PNG model
\cite{Prahofer2000} (marked by full horizontal lines).
}
\end{figure}

With a non-symmetric term being present there is no longer any reason for the height
distribution $P_L(h,t)$ to be symmetric with respect to $h$. Although in 1+1 
dimensions a KPZ interface of a {\it finite} system after saturation still happens to be
symmetric and Gaussian (see e.g.~\cite{Barabasi}), the profile of a {\em roughening} 
KPZ interface before saturation is indeed skewed~\cite{Zhang}, reflecting the asymmetry of 
the nonlinear term. In what follows we therefore restrict ourselves
to the roughening process before saturation, i.e., $t \ll L^z$, regarding a virtually
infinite system at finite times. Formally speaking, this can be achieved by
taking the thermodynamic limit $L \to \infty$ {\it before} the
time--asymptotic limit $t \to \infty$ is carried out. 
In particular, we are interested in the scaling function
$f(z) := f(z,0)$, which renders the rescaled shape of the skewed profile after 
sufficiently long time (see left panel of Fig.~\ref{figprofile}).

The function $f(z)$ is known to be universal, i.e.,
the asymptotic shape of the skewed profile is 
fully determined by the underlying KPZ field theory 
and does not depend on the microscopic details of the model. 
It has been suggested that the finite--time height distribution,
especially the form of its tails, is
approximately given by a stretched exponential
\beq
P_L(h,t) \propto \exp \left[-\mu\left(|h-\langle h
    \rangle|/t^{\gamma}\right)^{\eta_{\pm}}\right] \quad t \ll L^z \,,
\label{SkewP}
\eeq
meaning that $f(z) \sim \exp(-\mu|z|^{\eta_\pm})$. Here $\mu$ is a 
metric factor while the exponents 
$\eta_{\pm}$ refer to the two different tails of the distribution with
$\pm \lambda (h-\langle h \rangle)> 0$.
Because of the skewness both exponents are expected to be different.
An argument based on a replica scaling
analysis of directed polymers~\cite{Zhang2}, whose free energy fluctuations
corresponds to the height fluctuations of a KPZ interface, 
suggests the value $\eta_+ = 3/2$.

As a breakthrough, Pr{\"a}hofer and Spohn have shown recently~\cite{Prahofer2000} 
that the finite--time rescaled height
profile of the polynuclear growth model (PNG) \cite{PNG}, a model which is believed 
to belong to the KPZ universality class, equals the Gaussian orthogonal ensemble
(GOE) Tracy-Widom distribution. This immediatly leads to 
\beq
\eta_+ = 3/2 \,, \quad \eta_-=3.
\eeq
Numerical simulations reported in literature 
concerning both directed polymers and KPZ lattice models~\cite{KBM91a} 
give $\eta_+ = 1.6(2)$ and $\eta_- = 2.4(2)$, the latter value being not in
agreement with theoretical predictions. However, since these results
were obtained more than a decade ago the numerical precision was limited.
Performing similar simulations using the so-called single 
step model~\cite{SSM} (see Sec. 2) we measured the effective exponent 
$
\eta_{\rm eff}(z) = \frac{z}{\ln f(z)} \, \frac{d}{dz} \ln f(z)
$
which according to Eq.~(\ref{SkewP}) should converge to $\eta_\pm$ as $|z| \to \infty$. 
As it can be seen in the right panel of Fig.~\ref{figprofile}, our numerical results 
are in agreement with theoretical predictions for the PNG model while being incompatible with the previous estimate for $\eta_-$ of Ref.~\cite{KBM91a}.
Clearly, the center of the scaling function (i.e. small values of $z$) is
described only approximately by Eq.~(\ref{SkewP}).

\begin{figure}
\centerline{\includegraphics[width=80mm]{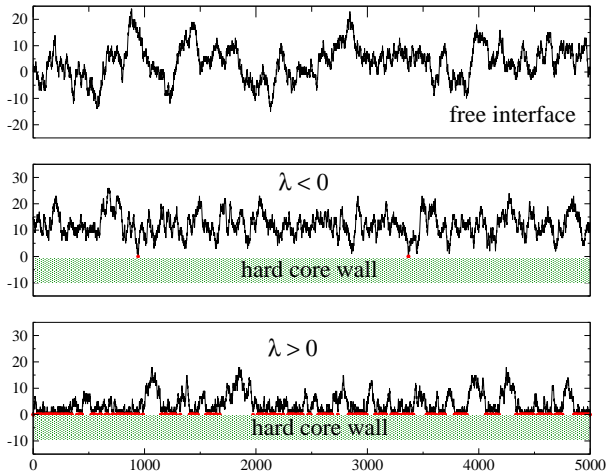}}
\caption{\label{figwall} \small
Snapshots of a roughening interface in a reference frame where the asymptotic
velocity is zero. The figure shows a free interface with $\lambda{<}0$ (upper panel)
compared to interfaces confined by a co-moving lower wall in the two cases $\lambda{<}0$
(middle panel) and $\lambda{>}0$ (lower panel). 
Simulations have been performed using the single step model (see Sec. 2) and
snapshots have been taken after 2048 time steps. }
\end{figure}

Looking at a snapshot of a roughening interface in 1+1 dimensions, it is almost
impossible to recognize the influence of the KPZ nonlinearity by naked eye.
Its influence, however, is much more pronounced in the presence of a hard-core wall.
The wall is fixed in a frame where the asymptotic velocity of the interface vanishes
and interacts with the interface solely by preventing excursions to negative heights. 
As can be seen in Fig.~\ref{figwall}, in presence of such a wall one can easily 
appreciate the dramatic difference emerging when the sign of $\lambda$ is 
changed\footnote{Alternatively one may compare a lower and an upper wall while
keeping the sign of $\lambda$ fixed.}. 
Surprisingly, for $\lambda<0$ the interface touches the wall only occasionally, 
while a high density of contact points is observed for $\lambda>0$.

The properties of a KPZ interface close to a reflecting wall has been studied recently 
in the context of nonequilibrium wetting~\cite{Wetting}, where the interface
describes a wetting layer on a planar substrate. Upon varying the average
growth velocity $v_0$ the interface undergoes a {\it depinning} transition between a 
pinned phase, in which portions of the interface remains attached to the wall, to a
depinned phase, where the interfaces detaches entirely and starts moving upwards.
At the critical point, where the asymptotic interface velocity is zero,
a second order phase transition takes place and various scaling laws can be singled out.
In addition, the case $\lambda <0$ describes the critical properties of most synchronization 
transitions in spatially extended chaotic systems \cite{Livi, Pikov}.

Previous numerical simulations suggested that the temporal 
decay of  the density $\rho_0(t)$ of contact points, where the interface touches the wall, 
obeys the power law ~\cite{MN1,MN2}
\begin{equation}
\label{theta}
\rho_0(t) \sim t^{-\theta},\qquad \theta \approx
\left\{
\begin{array}{cc} 
1.1(1) & \text{  if } \lambda<0 \\
3/4  & \text{  if } \lambda=0 \\
0.22(2) & \text{  if } \lambda>0 
\end{array}
\right.
\quad,
\end{equation}
where the exponent $3/4$ can be obtained from a transfer matrix calculation
\cite{Wetting}. Moreover, a hyperscaling relation observed in simulations 
starting with a single pinned site~\cite{Droz} 
suggests the rational value $\theta = 7/6$. 
Obviously, the different values of the exponents reflect the asymmetry of the
nonlinear term with respect to reflections $h \to -h$. Moreover, the pronounced numerical
variation of $\theta$ by a factor of 5 explains why the snapshots 
in Fig.~\ref{figwall} are so strikingly different.

Interestingly, the profile of a 
roughening KPZ interface next to a wall cannot be described
in terms of an appropriate generalized GOE Tracy-Widom distribution because of
emerging nonlinear term. More generally, the critical behavior of such a bounded growth process in
1+1 dimensions is not easily accessible by analytical
means. For example, renormalization group techniques fail either
due to the presence of a strong--coupling fixed point unaccessible 
by perturbative approaches in the case $\lambda<0$~\cite{MNRG} or 
due to essential singularities arising for $\lambda>0$.
Therefore, the primary aim of the present paper is to
discuss this case within a suitable mean field approximation.
The mean field theory to be constructed should incorporate the asymmetry caused by
the nonlinear term and should render a skewed height distribution with similar
properties as in the full model. Although the mean field theory ignores space,
it should not resemble a naive infinite-dimensional limit (where a KPZ interface 
is always smooth), instead it should reflect to some extent the dimensionality of
space in the thermodynamic limit $L \to \infty$. Moreover, the desired theory 
should be as simple as possible and exactly solvable. In the following
sections we propose and solve a mean field theory which meets these requirements. 

\section{Mean field equations}

\begin{figure}
\centerline{\includegraphics[width=125mm]{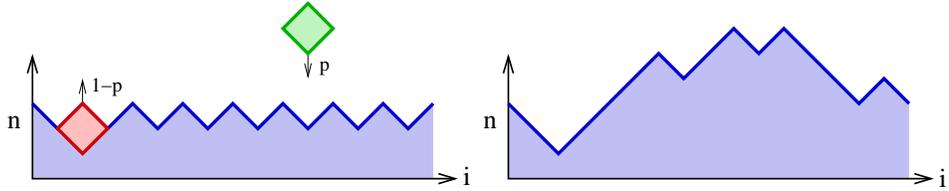}}
\caption{\label{figssm} \small
Single step model in 1+1 dimension. Left panel: The simulation usually starts with
a flat interface, realized as a horizontal sawtooth pattern.
On selecting a site with a local minimum
a diamond (rhombus) is deposited with probability~$p$, flipping up the interface
by two units. Similarly, if the selected site happens to be a local maximum, a diamond 
may evaporate with probability $1-p$, flipping the interface downward by two units.
Right panel: For $p>\frac12$ the interface roughens and propagates upwards.
}
\end{figure}

The mean field equations proposed here are inspired by a particular model,
the so-called single step model (SSM)~\cite{SSM}, which is probably the simplest 
and most compelling lattice model for KPZ-type interface growth.

In the single step model the growing interface is represented by a set of 
integer heights $n_i \in \mathbbm{N}$ residing at the sites $i=1\ldots L$ of a 
one-dimensional lattice of length $L$ with periodic boundary conditions, 
obeying the restriction
\begin{equation}
\label{restriction}
n_{i+1}-n_i=\pm 1\,.
\end{equation}
The interface evolves in time according random sequential updates as follows: 
At each sub--time step $dt=1/L$ a site $i$ is chosen at random. 
If the interface has a local minimum at site $i$ (i.e., $n_i < n_{i\pm 1}$) 
the height $n_i$ is increased by $2$ with probability $p\in[0,1]$. This 
update can be pictured as depositing a diamond (see Fig.~\ref{figssm}), 
transforming a local minimum into a local maximum. Similarly, if the selected site 
happens to be a local maximum ($n_i > n_{i\pm 1}$) the height $n_i$ is decreased 
by two units with probability $1-p$.

For $p=1/2$ the propagation velocity of the interface is zero and
the evolution rules satisfy detailed balance, as described by the 
EW equation. For $p \neq 1/2$, however, the propagation velocity
is nonzero, depending on the roughness and the average slope of the interface.
In this case the SSM exhibits KPZ growth with 
$\lambda$ being equal to $\frac12-p$.

By identifying upward segments $n_{i+1}-n_i=1$ with particles and 
downward segments $n_{i+1}-n_i=-1$ with vacancies, 
the single step model can be mapped exactly onto a partially
asymmetric exclusion process (ASEP)~\cite{ASEP} of diffusing particles with density $1/2$. 
Since the ASEP is known to evolve towards an uncorrelated product state with a
current $j=p/2-1/4$, it is immediately clear that the propagation velocity of
the interface tends to
\begin{equation}
v_\infty = \lim_{t \to \infty} v(t) = p-\frac12
\end{equation}
as $t \to \infty$.\footnote{Initially the velocity is higher, and the 
excess velocity $|v(t)-v_\infty|$ decays as $t^{-1/3}$.} 
Moreover, the mapping to the ASEP allows one to solve the 
model via Bethe ansatz\cite{Gwa92}, making it possible to derive the
KPZ dynamical exponent $z=3/2$ and various other quantities exactly. 
Other rigorous results concerning shape fluctuations in the ASEP can also be
found in Ref. \cite{Johansson}.

In order to formulate a mean field theory for the single step model,
let $N_u(n,t)$ and $N_d(n,t)$ be the probabilities of finding an upward or downward 
segment with their lower edge rooted at height level $n$. Let us first consider a deposition 
process, in which a local minimum at level $n$ is flipped into a local maximum
at level $n+2$.
Having selected a random site, the probability to find such a local minimum
at a given height can
be approximated as follows. Clearly, the probability of finding a downward
segment on the left side terminating at height $n$ is $N_d(n,t)/L$, where $L$
is the system size. With this probability, knowing that the height of the selected site
is $n$, the adjacent segment to the right can only go up or down so that
the \textit{conditional} probability to find an upward segment is
given by $N_u(n,t) /(N_u(n,t)+N_d(n-1,t))$. Ignoring possible correlations
the total probability of finding a local minimum at 
height $n$ is the product of these two expressions.
The deposition process, taking place with probability $p$, therefore leads to a loss
of probability at level $n$
\begin{equation}
\label{depositionevent}
\begin{split}
N_u(n,t) \to N_u(n,t+dt) &= N_u(n,t) - \frac{p}{L} \, \frac{N_d(n,t)N_u(n,t)}{(N_u(n,t)+N_d(n-1,t))}\\
N_d(n,t) \to N_d(n,t+dt) &= N_d(n,t) - \frac{p}{L} \, \frac{N_d(n,t)N_u(n,t)}{(N_u(n,t)+N_d(n-1,t))}
\end{split}
\end{equation}
and a corresponding gain at level $n+1$. Similar expressions can be derived 
for the evaporation process. Obviously, this approximation accounts 
for the restriction~(\ref{restriction}) and the one-dimensional structure 
of the model but disregards possible nearest-neighbor correlations. 

The structure of Eq.~(\ref{depositionevent}) suggests that the probabilities $N_u(n,t)$ and $N_d(n,t)$ evolve exactly in the same way. In fact, it is easy to see that
in a system with periodic boundary conditions the numbers of
upward and downward segments are exactly equal. Assuming the same to hold
in an infinite system we have
\begin{equation}
N_u(n,t)=N_d(n,t)
\end{equation}
for every $n$ and $t$. Thus, introducing a combined probability density
\begin{equation}
P(n,t) = \frac{N_d(n,t)+N_u(n,t)}{2L}
\end{equation}
the loss at level $n$ due to a deposition event in Eq.~(\ref{depositionevent}) can be recast as
\begin{equation}
P(n,t) \to P(n,t+dt)=P(n,t)-p\,\frac{P_n(t)^2}{P_n(t)+P_{n-1}(t)} \,.
\label{mf2}
\end{equation}
Collecting all loss and gain contributions due to deposition and evaporation
one arrives at the following set of mean field equations
\begin{eqnarray}
\frac{\partial}{\partial t} P_n(t) \;=\;
p\left[ \frac{P_{n-1}(t)^2}{P_{n-1}(t)+P_{n-2}(t)} -
\frac{P_n(t)^2}{P_n(t)+P_{n-1}(t)}\right] \nonumber \\
+(1-p)\left[ \frac{P_{n+1}(t)^2}{P_{n+1}(t)+P_{n+2}(t)} -
\frac{P_n(t)^2}{P_{n}(t)+P_{n+1}(t)}\right] 
\label{MF}
\end{eqnarray}
which serve as a starting point for all further calculations throughout this paper.
Notice that the form of the denominators appearing on the r.h.s of Eq.~\toref{MF}
is a consequence of the restriction~(\ref{restriction}), and that some care 
has to be taken when one of them vanishes. Since the numerators are quadratic
we assume that each of these terms is zero {\it whenever} 
their denominator vanishes.

Introducing a probability current flowing between neighboring levels
\begin{equation}
\label{current}
J_{n,n+1}(t) = p\,\frac{P_n(t)^2}{P_n(t)+P_{n-1}(t)} \,-\, 
(1-p)\frac{P_{n+1}(t)^2}{P_{n+1}(t)+P_{n+2}(t)}
\end{equation}
these equations can be also written as
\begin{equation}
\label{MFJ}
\frac{\partial}{\partial t} P_n(t) \;=\;
J_{n-1,n}(t) - J_{n,n+1}(t) \,.
\end{equation}
Obviously, they conserve probability $\sum_{k=-\infty}^{+\infty} P_k(t)$
so that the integrated probability distribution
\begin{equation}
Q_n(t) := \sum_{k=-\infty}^{n} P_k(t) 
\end{equation}
satisfies the simple evolution equation
\begin{equation}
\frac{\partial}{\partial t} Q_n(t) \;=\; -J_{n,n+1}(t) \,.
\end{equation}
By construction these mean field equations reflect both the one-dimensional
structure as well as the restriction but they ignore spatial correlations between 
the segments. The full model does exhibit such correlations, but it evolves towards 
a trivial state without correlations (corresponding to a simple
product state in the ASEP). Although this trivial state is never reached
in an infinite system, it may explain why the mean field equations proposed
here reproduce so many of the observed phenomena faithfully, some of them
even exactly, as will be shown in the following sections.

\section{Exact solution of the mean field equations}

Let us first consider the case of a free interface, where the height index $n$
runs over all integers from $-\infty$ to $+\infty$. 

As the KPZ equation is invariant under appropriate rescaling of 
space, time and height \cite{Barabasi}, we can carry out the continuum limit by
introducing a new height variable $h=n\Delta$, where $\Delta$ is the new
height unit of the rescaled system. 
In order to investigate the asymptotic properties of the roughening processes let us assume that
$P_n(t)$ varies only slowly with $n$
and expand the r.h.s. of Eq.~(\ref{MF}) 
as a Taylor series around $h$. Keeping contributions up to fourth order 
in $\Delta$ we obtain the partial differential equation
\begin{eqnarray}
\label{PDMF}
\frac{\partial}{\partial t} P(h,t)&=&
\frac{\Delta(1-2p)}{2} P'(h,t) + \\
&&\frac{\Delta^3(2p-1)}{24}\left[\frac{3P'(h,t)^3}{P(h,t)^2}-
                                  \frac{6P'(h,t)P''(h,t)}{P(h,t)}+
                                  4P'''(h,t)\right] +
\nonumber\\
&&\frac{\Delta^4}{8}\,\,\biggl[\frac{2P'(h,t)^4}{P(h,t)^3}-
\frac{5P'(h,t)^2P''(h,t)}{P(h,t)^2}+ \nonumber\\
&& \hspace{9mm}
\frac{2P''(h,t)^2}{P(h,t)}+\frac{2P'(h,t)P'''(h,t)}{P(h,t)}-P''''(h,t)\biggr] \nonumber\,,
\end{eqnarray}
where the prime stands for a partial derivative with respect to $h$. Obviously
the leading term 
of order $\Delta$ on the r.h.s. generates a uniform propagation of the
probability distribution, hence, the average height
$\langle h(t) \rangle$ of the interface will asymptotically grow with the 
linear velocity
\beq
\label{vel}
v=\Delta\left(p-\frac12\right)
\eeq
plus some sublinear correction terms.
Assuming ordinary Family-Vicsek scaling~\cite{FVScaling}, 
it is therefore near at hand to test the validity of the scaling form
\begin{equation}
\label{ansatz}
P(h,t) \;=\; t^{-\gamma} \, f\Bigl( \frac{h-vt}{t^{\gamma}}\Bigr)
\end{equation}
which -- by definition -- conserves the integrated probability
$\int_{-\infty}^{+\infty} \text{d}h P(h,t)$. 
Notice that the normalization of the height probability distribution
implies the scaling function $f(z)$ to be normalized as well.
In what follows we solve Eq.~\toref{PDMF} both in the equilibrium case
$p=1/2$ and the non--equilibrium case $p\neq 1/2$ confirming that
our results do not depend on $\Delta$. In particular, we will show that
higher order terms appearing in the expansion of Eq. \toref{MF}
turns to be irrelevant, vanishing in the asymptotic limit $t \to \infty$.
The correct asymptotic behavior of Eq. \toref{MF} will be therefore recovered by setting
$\Delta =1$.

\subsection{Equilibrium roughening of a free interface}
%
%
We start analyzing the special case $p=1/2$, where the dynamic processes of 
the full model are known to exhibit detailed balance. In this case the velocity
$v$ is zero and the first-order and third-order contributions 
on the r.h.s. of Eq.~\toref{PDMF} vanish. 
Inserting the Ansatz~\toref{ansatz} into Eq.~(\ref{PDMF}) we find that, 
up to fourth order, the partial differential equation reduces to a non-trivial 
ordinary differential equation for the scaling function 
if and only if $\gamma_{\rm eq}=1/4$ (the subscript denoting the equilibrium case).
This is exactly the value predicted by the EW theory for equilibrium roughening.
Moreover one easily notices that by fixing $\gamma_{\rm eq}=1/4$
higher order terms $O(\Delta^5)$ occurring in the Taylor expansion 
of Eq.~\toref{MF} are irrelevant in the 
asymptotic limit $t \to \infty$. The differential equation therefore reads
\begin{eqnarray}
\label{EW1}
&&\frac{1}{f_{\rm eq}(z)^3}\,\biggl[f_{\rm eq}(z)^4 -\frac{5}{2}\Delta^4 f_{\rm eq}(z)f_{\rm eq}'(z)^2f_{\rm eq}''(z)+\\
&&\hspace{15mm}\Delta^4 f_{\rm eq}(z)^2 \Bigl(f_{\rm eq}''(z)^2 +f_{\rm eq}'(z)f_{\rm eq}'''(z)\Bigr) +
\nonumber \\
&&\hspace{15mm}f_{\rm eq}(z)^3\Bigl(zf_{\rm eq}'(z)-\frac{\Delta^4}{2}f_{\rm eq}''''(z)\Bigr) +
\Delta^4 f_{\rm eq}'(z)^4 \biggr] = 0\,,\nonumber
\end{eqnarray}
where $z=h\,t^{-\gamma_{\rm eq}}$ denotes the scaling variable. Integrating both sides we obtain
\begin{equation}
\label{EW2}
zf_{\rm eq}(z) - \frac{\Delta^4}{2} \left[\frac{f'_{\rm eq}(z)^3} {f_{\rm eq}(z)^2} 
- 2 \frac{f'_{\rm eq}(z) f''_{\rm eq}(z)}{f_{\rm eq}(z)} 
+ f'''_{\rm eq}(z)\right] = 0\,.
\end{equation}
This equation admits the two simple solutions
\beq
f^{\rm free}_{\rm eq}(z) = 
\frac{1}{\Delta 2^{1/4} \sqrt{\pi}} \exp \left(-\frac{z^2}{\Delta^2 \sqrt{2}} \right)
\label{EW2free}
\eeq
and
\beq
f^{\rm bound}_{\rm eq}(z) = 
\frac{2^{1/4}}{\Delta^3 \sqrt{\pi}} \,z^2\,\exp \left(-\frac{z^2}{\Delta^2 \sqrt{2}} \right)
\label{EW2bound}
\eeq
which have been normalized over the real line. 
The first solution $f^{\rm free}_{\rm eq}$ is a simple Gaussian and
represents the physical solution for a free interface 
starting with a flat initial condition $h(x,t)=0$. 
The second solution $f^{\rm bound}_{\rm eq}$ is characterized by two different
maxima over the real line and is therefore dismissed as unphysical
in the free case. However, as we will see in Sect. 4, this solution becomes physically
meaningful in the presence of a hard-core wall at zero height.

To summarize we note that the mean field equation for $p=1/2$ does indeed capture 
the features of one-dimensional EW roughening 
in the thermodynamic limit $L\to\infty$.

\subsection{Nonequilibrium roughening of a free interface}

We now turn to the nonequilibrium case $p\neq1/2$. Inserting again the 
scaling form~\toref{ansatz} and the expression for the velocity~\toref{vel} 
into the partial differential equation~\toref{PDMF}, we find that, 
up to {\em third} order, the partial differential equation reduces to a 
non-trivial ordinary differential equation for the scaling function 
(i.e., without explicit occurance of $t$)
if and only if $\gamma=1/3$. The differential equation then reads
\begin{eqnarray}
\label{NonlinDE}
&&\frac{1}{f(z)^2} \Bigl[8 f(z)^3 + 3 k f'(z)^3 - 6 k f(z)f'(z)f''(z) + \\
&& \hspace{12mm} 4 f(z)^2 \bigl(2z\,f'(z) + k f'''(z)\bigr)\Bigr] \;=\; 0\,,\nonumber
\end{eqnarray}
where 
$
z =  (h-vt)\,t^{-\gamma}
$
denotes the scaling variable in the comoving reference frame and 
\begin{equation}
k=(2p-1)\Delta^3\neq 0\,.
\label{metricf}
\end{equation}
As in the equilibrium case the postulate of Family-Vicsek scaling 
applied to the mean field equation already determines the roughening exponent. 
Remarkably the value $\gamma=1/3$ coincides exactly with the known value 
of a KPZ process in 1+1 dimensions.

As can be verified easily, upon fixing $\gamma=1/3$, 
the fourth order terms (and all higher order terms) 
of the Taylor expansion turn out to be irrelevant in the asymptotic limit 
$t \to \infty$ and thus generate only short time corrections to the scaling function.
It is also worth commenting that asymptotic EW scaling can only be seen in the 
symmetric case $p=1/2$. For any small deviation from this value the third-order
terms do not vanish, leading eventually to a crossover to KPZ scaling in the
limit $t \to \infty$. Therefore, the mean field equations
nicely reproduce the character of the KPZ nonlinearity as a relevant perturbation.

Assuming that $f(z) \neq 0$ and integrating both sides 
of Eq.~(\ref{NonlinDE}) one obtains a simplified equation
%
%
%
%
which, by substituting $f(z)=u(z)^4$, can be further reduced to a simple Airy differential equation
\begin{equation}
zu(z)+2ku''(z)\;=\;0 
\label{Airy}
\end{equation}
with the general solution
\begin{equation}
\label{generalsolution}
u(z) = \left\{
\begin{array}{cc}
 c_1 \text{Ai}\Bigl(\frac{-z}{(2k)^{1/3}}\Bigr) +
 c_2 \text{Bi}\Bigl(\frac{-z}{(2k)^{1/3}}\Bigr)
&
\quad \text{ for } k \geq 0 \\[3mm]
 c_1 \text{Ai}\Bigl(\frac{z}{(-2k)^{1/3}}\Bigr) + 
 c_2 \text{Bi}\Bigl(\frac{z}{(-2k)^{1/3}}\Bigr)
&\quad \mbox{ for } k < 0
\end{array}
\right.
\end{equation}
where $\text{Ai}(z)$ and $\text{Bi}(z)$ are Airy functions (see for
instance \cite{Abramowitz}).
For given $|k|$ the two solutions differ only by a reflection $z\to -z$ so
that for the rest of this section we can
restrict ourselves to the case $k>0$ (which corresponds to a negative
nonlinear term, i.e. $\lambda < 0$). 

The two integration constants have to be chosen such that $f(z)$ is properly normalized
and the appropriate boundary conditions are satisfied. For a free interface,
the scaling function $f(z)$ 
has to vanish for $z\to \pm \infty$ in such a way that all of its moments are
finite, hence $c_2=0$.  Surprisingly, the remaining solution
{\em oscillates} for $z \to \infty$ and does not vanish fast enough to yield
finite moments.
We conclude that the physically meaningful 
solution extends from $z=-\infty$ to the first root of the Airy function $z_0$ 
(where $f(z_0)=f'(z_0)=f''(z_0)=0$) and
vanishes elsewhere. The solution for the free interface therefore reads
\begin{equation}
\label{freesolution}
f(z) \;=\; 
\left\{
\begin{array}{ll}
\frac{1}{\mathcal N}\,\left[\text{Ai}\Bigl(\frac{-z}{(2k)^{1/3}}\Bigr) \right]^4
&\quad \text{ for } -\infty < z < z_0 \\[3mm]
0 & \quad \text{ for } \,\, z_0 \leq z < +\infty 
\end{array}
\right.
\end{equation}
where $z_0 \simeq 2.94583 \, k^{1/3}$ and ${\mathcal N} \simeq 0.127153 \,
k^{1/3}$. As it can
be seen, the parameter $k$, describing the strength of the KPZ nonlinearity,
appears here as a simple metric factor in the scaling function. Notice that
by Eq. \toref{metricf} the height unit $\Delta$ has been absorbed in $k$. 

\begin{figure}
\centerline{\includegraphics[width=75mm]{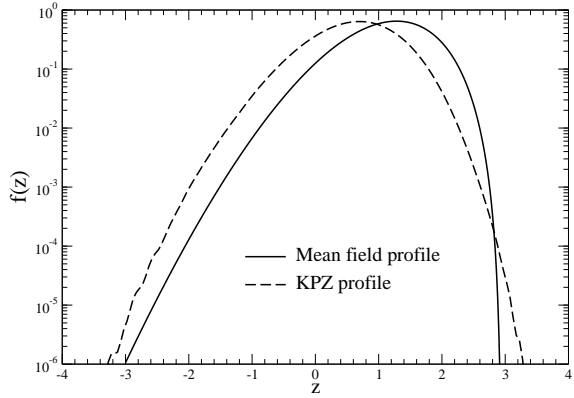}}
\caption{\label{figfreeprofile} \small
Rescaled skewed profile obtained within the mean field approximation
(solid line) compared to the numerically determined KPZ profile 
using the single step model with $p=1$. 
Differences between the two profiles can be highlighted by 
their mean value $a_1$. The mean value controls the KPZ excess velocity and 
it is known to scale as $\lambda^{1/3}$.
While the mean field profile is characterized by $a_1 \simeq 1.13184$,
direct numerical estimate renders $a_{1,SSM} \simeq 0.60$.
The vertical axis is plotted in a logarithmic scale.
}
\end{figure}

Figure~\ref{figfreeprofile} shows the solution~(\ref{freesolution}) in comparison with
the numerically determined profile of a freely roughening KPZ interface. Although the 
two curves are different due to the approximative character of the mean field theory,
they share essential qualitative features. To quantify those it is instructive to
compute the skewness 
\begin{equation}
S = \frac{c_3}{c_2^{3/2}} = \frac{a_3-3a_2a_1+2a_1^3}{(a_2-a_1^2)^{3/2}} \quad,
\end{equation}
where $c_n$ is the $n^{\rm th}$ moment of the height probability distribution and
$a_n=\int_{-\infty}^{z_0} dz\,z^n f(z)$ denotes the $n^{\rm th}$ central 
moment of the rescaled
profile $f(z)$. The skewness is expected to be
universal in modulo for all KPZ growth processes, with 
its sign being equal  
to the sign of the nonlinear term.
Known numerical estimates \cite{KrMHH92} give the value
$|S_{\rm KPZ} | = 0.28 \pm 0.04$,
which is in good agreement with the theoretically computed skewness for the
PNG model\footnote{The PNG model is characterized by a positive nonlinear term.} 
\cite{Prahofer2000}, 
\beq
S_{PNG} \simeq 0.2935 \ .
\eeq
Mean field theory, on the other hand, renders the value
\begin{equation}
S_{\rm MF} \simeq \pm 0.465970 \,, 
\end{equation}
where the positive (negative) sign correspond to the case $k<0$ ($k>0$).
Although this value is different from direct numerical estimates,
it has the correct sign and the same order of magnitude, showing that the mean field
theory captures qualitatively the influence of the nonlinear KPZ term.

Surprisingly, the mean field theory predicts that the interface profile is
asymptotically bounded for $z>0$ at a {\it finite} value $z_0$. This means that 
within mean field the advancing front of the distribution exhibits a sharp
cutoff rather than a stretched exponential tail. However, on the opposite side,
where $z$ is negative, the profile does indeed decay as a stretched exponential:
\begin{equation}
f(z) \sim |z|^{-1/4} \, e^{-\frac{2}{3\sqrt{2k}} |z|^{3/2}}
\qquad (z \to -\infty)
\end{equation}
This result suggests that $\eta_+=3/2$, which coincides with the theoretical 
value predicted in the context of directed polymers and of the PNG model. 
On the other hand, $\eta_-$ does not exist within the 
mean field approximation, which therefore fails to correctly describe large
negative (w.r.t. the sign of the KPZ nonlinearity) height fluctuations. 

\section{Roughening in the vicinity of a wall}
%
%
We now modify the single step model and the associated 
mean field equations in order to incorporate the presence of
a hard core wall. Our aim is to determine 
the surface exponent $\theta$ introduced in
Eq.~(\ref{theta}) within the mean field approximation. 
In terms of the continuous height variable $h$,
the density of pinned sites can be defined as the integral
of the height probability distribution between the hard core wall and some
arbitrary small height level $h_0$, i.e., 
\beq
\rho(t) = \int_{vt }^{h_0+ vt } P(h,t)\,dh 
=  \int_{0}^{h_0 t^{-\gamma}}
f(z) \,dz \quad,
\label{pippo}
\eeq
so that the surface exponent is completely determined by the behavior of the
scaling function $f(z)$ for $z\ll 1$.

In order to formulate the appropriate boundary condition in the
continuum limit, one has to resort to the discrete formulation of the problem.
Following the approach outlined in Ref. \cite{Ginelli03},
the wall is initially located at zero height 
and moves {\it discontinuously}
with the average velocity $v=p-\frac12$, its actual height being given by
$n_0(t)=\lfloor vt \rfloor$ (where $\lfloor \cdot \rfloor$
indicates the integer part).
The interface is restricted to evolve above
the wall, i.e.
\begin{equation}
P_n(t)=0 \qquad (n \leq n_0(t))
\end{equation}
so that the mean field equations \toref{MF} have to be modified accordingly
close to the wall. In particular there is no probability
current between level $n_0$ and $n_0+1$, so that
\begin{equation}
\label{MFJbnd}
\frac{\partial}{\partial t} P_{n_0+1}(t) \;=\; - J_{n_0+1,n_0+2}(t) \quad,
\end{equation}
while Eq. \toref{MFJ} still holds for $n > n_0+1$.
Depending on $p$ one has to distinguish three different cases.
If $v>0$ the wall advances by one unit in time intervals $\Delta t = 1/v$, flipping
up all local minima at level $n_0+1$ by two units. This means that $P_{n_0+1}$ is
increased by $P_{n_0}$ while $P_{n_0}$ is set zero. If $v<0$ the wall retracts
by one unit in time intervals $\Delta t = |1/v|$, allowing height level $n_0$, which
was previously set to zero, to become nonzero during the subsequent evolution.
Finally, for $v=0$ the wall does not move, i.e. $n_0=0$ for all times $t$.
The moving wall makes it difficult to specify the correct boundary conditions,
so out of equilibrium we will derive them in the special cases $p=1$ and 
$p=0$, where the KPZ nonlinearity is maximal. Our reasoning, which once more
relies on a series expansion in the proximities of the wall, shows that
both a pushing ($p=1$) and a retracting ($p=\frac12$ wall impose a Dirichlet
boundary condition for the scaling function. Surprisingly, it turns out that 
a retracting wall ($p=0$) does not fix any boundary condition for the scaling
function, which is free to assume any finite value at wall level, thus 
justifying the high density of pinned sites which is numerically observed
in the case $\lambda>0$ (see Fig. \ref{figwall}).
General scaling arguments suggest that results obtained for $p=1$ ($p=0$) hold for any
$p>1/2$ ($p<1/2$).

\subsection{Depinning in the equilibrium case $p=1/2$}
\label{EWdepin}

In the equilibrium case we have $v=0$ so that the wall does not move.
As shown in the Appendix, a wall at zero height 
imposes a Dirichlet boundary condition $f(0)=0$.
Obviously, the only solution satisfying this boundary condition is Eq.~\toref{EW2bound}
\beq
f^{\rm bound}_{\rm eq}(z) = 
\frac{2^{5/4}}{\Delta^3 \sqrt{\pi}} \,z^2\,\exp \left(-\frac{z^2}{\Delta^2 \sqrt{2}} \right)
\label{EW2boundnormalized}
\eeq
which has been normalized here over the positive real axis. With this solution
we can immediately read off the surface exponent from Eq. \toref{pippo},
\begin{equation}
\theta^{\rm MF}_{\rm EW} = 3/4.
\end{equation}
We note that this value coincides exactly with the known exponent for 
Edwards Wilkinson growth next to a wall.
%
\subsection{Depinning in the non-equilibrium case $p=1$}
\label{KPZdepin}
%
%
\begin{figure}
\centerline{\includegraphics[width=120mm]{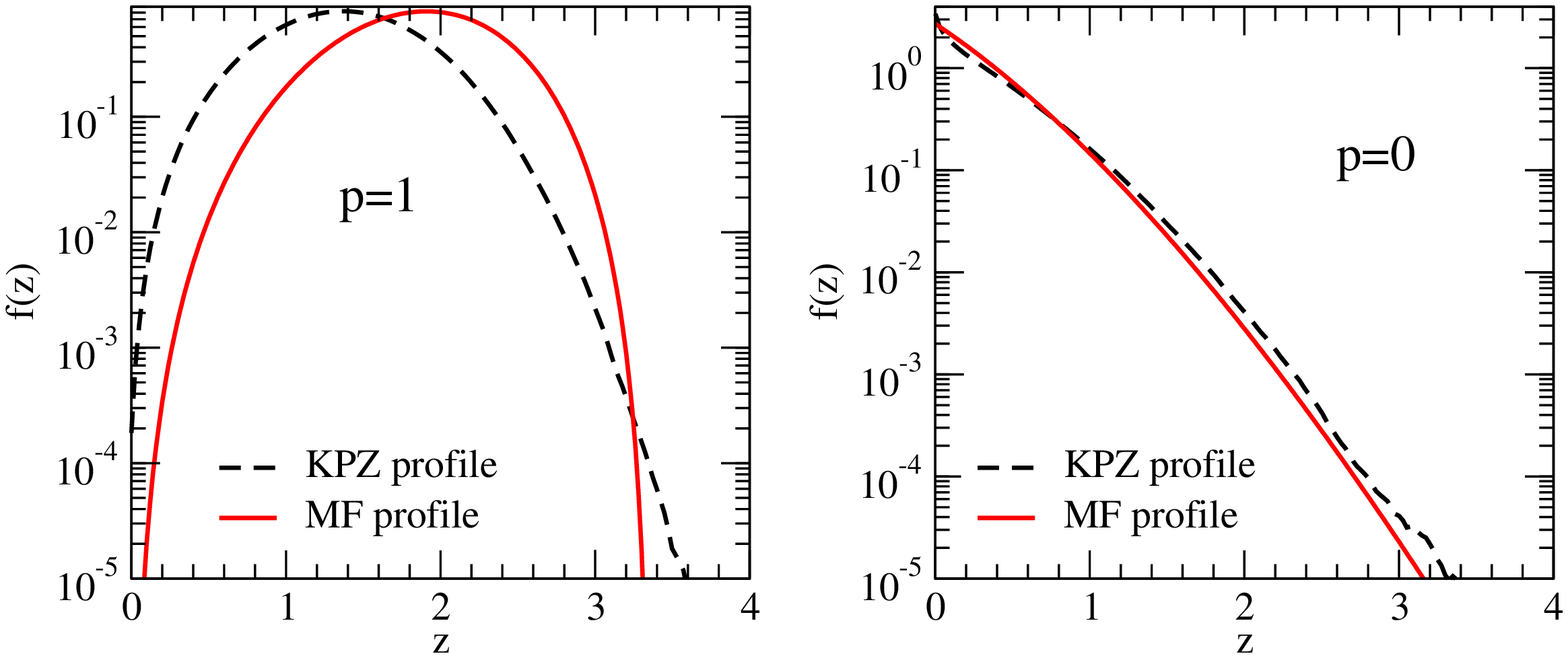}}
\caption{\label{figwithwall} \small
Rescaled mean field profiles in the presence of a hard core wall compared
to the corresponding profiles in the SSM for $p=1$ (left panel)
and $p=0$ (right panel). 
The vertical axis have been plotted in a logarithmic scale.
}
\end{figure}
%
For $p>1/2$ the wall advances discontinuously which makes it more difficult to
specify the boundary conditions. As shown in the Appendix, in this case 
the co-moving wall again leads to a Dirichlet
boundary condition $f(0)=0$ for the scaling function. According to 
Eq.~(\ref{generalsolution}) the corresponding solution then reads
\begin{equation}
\label{uppersolution}
f(z) \;=\; 
\left\{
\begin{array}{ll}
\frac{1}{\mathcal N}\,\left[\text{Ai}\Bigl(\frac{-z}{(2k)^{1/3}}\Bigr) -
\frac{1}{\sqrt{3}} \text{Bi}\Bigl(\frac{-z}{(2k)^{1/3}}\Bigr)\right]^4
&\quad \text{ for } 0 \leq z < z_1 \\[3mm]
0 & \quad \text{ for } \,\, z_1 \leq z < +\infty 
\end{array}
\right.
\end{equation}
where $z_1 \simeq 3.32426 \, k^{1/3}$ is the first positive root of
the scaling function $f(z)$ (where also its first three derivatives vanish) 
and ${\mathcal N} \simeq 0.133454 \, k^{1/3}$.
As in the free case the profile exhibits a sharp cutoff (see Fig.~\ref{figwithwall}), 
although at a different value of $z$. Since $f(z) \sim z^4$ for $z \to 0$, 
the surface exponent is given by
\begin{equation}
\theta^{\rm MF}_{\rm p=1} = 4/3,
\end{equation}
This values has to be compared with the numerical estimate 
$\theta^{\rm KPZ}_{\lambda{<}0} = 1.1(1)$ in Eq.~(\ref{theta}).

\subsection{Depinning in the non-equilibrium case $p=0$}
%
For $p<1/2$ the wall moves discontinuously backward. 
As shown in the Appendix, this situation is special in so far as the retracting
wall does not specify any boundary condition on the scaling function, allowing $f(0)$
to be nonzero. In fact, according to Eq.~(\ref{generalsolution}) the only 
normalizable solution is given by
\begin{equation}
\label{lowersolution}
f(z) \;=\; 
\frac{1}{\mathcal N}\,\left[\text{Ai}\Bigl(\frac{z}{(2k)^{1/3}}\Bigr)\right]^4
\end{equation}
with the normalization constant ${\mathcal N} \simeq 0.00584355 \, k^{1/3}$.
Since $f(0)>0$ the surface exponent is given by
\begin{equation}
\theta^{\rm MF}_{\rm p=0} = 1/3,
\end{equation}
This values has to be compared with the numerical estimate 
$\theta^{\rm KPZ}_{\lambda{>}0} = 0.22(2)$ in Eq.~(\ref{theta}).

\section{Conclusions}
In this paper we presented a mean field theory for 1+1 dimensional nonlinear
growth processes evolving according to a KPZ equation. 
It is worth stressing that our approach does not neglect 
all types of fluctuations, as the term mean field usually suggests.
Instead our equations retain information about the one dimensional 
spatial structure and the height restriction $h_i - h_{i+1} = \pm 1$. 
Therefore, our mean field theory is not expected to hold
exactly above some upper critical dimension, rather it serves as an
approximation for the one dimensional case. Although we neglect 
spatial correlation between local slopes of 
the roughening interface, the theory has a very predictive power.
Its success can be ascribed 
to a fluctuation dissipation theorem which is known to hold only for the 1+1
dimensional case \cite{Barabasi}.
Moreover, the mean field equations presented herein have been derived 
in the thermodynamic limit $L \to \infty $, and thus they are suited for
probing finite time behavior, i.e. $t \ll L^z$. 

Our approach is therefore successful in predicting a power law decay for the
density of interfacial sites pinned to the substrate, thus supporting previous
numerical studies of the non--equilibrium case. Although the mean field 
surface exponents $\theta$ differ from the numerically estimated values, 
our theory correctly reproduces the dramatic difference between the surface 
behavior in the presence of a negative or positive nonlinear KPZ term. 

For what concerns the finite--time bulk properties of a KPZ interface, our
simple mean field theory correctly reproduces the exact value for the 
roughening exponent $\gamma=1/3$ and the skewed nature of the finite time
height probability distribution. We found that one of the two tails
of such a distribution decays as a stretched exponential with exponent 
$\eta_+ = 3/2$, thus confirming previous results obtained in the context 
of directed polymers and for the PNG model. 
While our theory successfully predicts large {\it positive} (w.r.t. to the sign of
the KPZ nonlinearity) height fluctuations, it fails in describing 
large {\it negative} ones, exhibiting a sharp cut--off for the negative tail.
It is therefore interesting to note that
a simple scaling argument \cite{PNG} proposed in the context  of the PNG model
directly relates $\eta_+$ to the roughening exponent, i.e. $\eta_+ =
1/(1-\gamma)$, while no corresponding argument can be worked out for
$\eta_-$. This is due to the fact that height fluctuations
with the same sign w.r.t. the KPZ nonlinear term manifest as ``bumps'' (or
``holes'') which grows laterally, while height fluctuations with the opposite
sign manifest as ``holes'' (or''bumps'') which shrink laterally. 

It is also worth noticing that the mean field theory reproduces correctly 
almost all bulk nonlinear critical properties at finite times, 
while it gives only approximate results for the surface exponent 
$\theta$. This is an indication that the substrate introduces 
spatial correlations between local slopes at low height levels.
It is our belief that a detailed study of the SSM with a hard substrate
may eventually lead to the exact analytical knowledge of critical
depinning properties. It would also be interesting to find out whether the
methods introduced in Ref.~\cite{Prahofer2000} 
can be applied to the problem of a KPZ interface with a wall.

Finally, equilibrium results are correctly reproduced as a marginal case, and
small out of equilibrium corrections eventually leads to full KPZ behavior.

\noindent
{\bf Acknowledgements}\\
We would like to thank A. Politi for stimulating discussions concerning the 
simplest formulation of a mean field equation for a KPZ interface.

%
%
\appendix
\section{Boundary conditions in the presence of a hard core wall}
\label{app}
%
In order to solve the mean field equation \toref{PDMF} when a hard core wall
is imposed, it is necessary to go back to the discrete formulation of 
the problem and to consider the 
proper evolution equation for the height probability distribution
\toref{MFJbnd} close to the wall. 
In the following we analyze the special cases $v=0$ and $v = \pm \frac12$ in
order to derive the corresponding boundary conditions for the scaling function $f(z)$.

\subsection{$p=\frac12, v=0$}

In this case the wall does not move and $n_0=0$ for all times $t$.
According to Eq.~(\ref{MFJbnd}) the density of contact points $P_1(t)$ then evolves as
\begin{equation}
\label{MFJ1}
\frac{\partial}{\partial t} P_1(t) \;=\; - J_{1,2}(t) \;,
\end{equation}
where 
\beq
J_{1,2}(t) = \frac{P_1(t)}{2}-\frac{P_2(t)^2}{2(P_2(t)+P_3(t))}=0.
\eeq
In the limit $t \to \infty$ this equation implies a Dirichlet boundary condition $f(0)=0$.
To see this let us assume that $f(0)\neq0$ with $|f'(0)|<\infty$. Assuming EW scaling
the first three probabilities would be given by 
$P_1(t) \simeq P_2(t) \simeq P_3(t) \simeq t^{-1/4} f(0)$ (where we expanded
the scaling function $f$ around $z=0$ keeping only the leading term),
giving rise to a current $J_{1,2}(t) \simeq \frac14 t^{-1/4} f(0)$. 
Since the l.h.s. scales as $t^{-5/4}$, Eq.~(\ref{MFJ1}) cannot hold unless $f(0)=0$. 

\subsection{$p=1, v=\frac12$}

For $p=1$ one has $\Delta t = 2$, i.e. 
the wall advances by one unit after every second time step.
To find the appropriate boundary condition for $f(z)$ (now assuming KPZ
scaling with $\gamma=1/3$), let us first consider 
the continuous temporal evolution between two advancements when the wall is
fixed at some height ${n_0}\lfloor -t_0/2 \rfloor$. 
As in the previous case the probabilities
$P_{n}(t)$ with $n \leq n_0$ vanish. According to Eqs. \toref{current}, \toref{MFJ} and 
\toref{MFJbnd}, the height probability distribution 
at the first two levels above the wall obeys to the differential equations
\begin{equation}
\label{LowerDGL}
\begin{split}
&\frac{\partial}{\partial t} P_{n_0+1}(t) \;=\; - P_{n_0+1}(t) \\
&\frac{\partial}{\partial t} P_{n_0+2}(t) \;=\; + P_{n_0+1}(t) - 
\frac{P_{n_0+2}^2} {P_{n_0+1}+P_{n_0+2}}\,.
\end{split}
\end{equation}
Let us again suppose that $f(0) \neq 0$ and $|f'(0)| < \infty$, i.e.,
just after the advancement of the wall at time $t_0$ we assume that 
to leading order $P_{n_0+1}(t_0) \simeq P_{n_0+2}(t_0) \simeq t_0^{-1/3} f(0) =: c\,(t_0)$.
Iterating the differential equations~(\ref{LowerDGL}) over two
time steps, and by assuming $c(t) \simeq c(t_0)$ for $t_0\leq t \leq t_0+2$,
one obtains $P_{n_0+1}(t_0+2)\simeq c\,(t_0) e^{-2}$.
On the other hand, by numerically solving the differential equation for level $n_0+2$ 
\beq
P_{n_0+2}(t) \;\simeq\; + c\,(t_0) e^{-(t-t_0)} - 
\frac{P_{n_0+2}^2} {c\,(t_0) e^{-(t-t_0)}+P_{n_0+2}}\,
\eeq
one gets
$P_{n_0+2}(t_0+2) \approx 0.596 \, c(t_0)$. At time $t_0+2$ the wall
advances by one unit, i.e. all local minima at height $n_0$ are flipped upwards, 
meaning that $P_{n_0+1}(t_0+2)$ is first added to $P_{n_0+2}(t_0+2)$ and then
set to zero. Just after advancement $P_{n_0+2}(t_0+2)\approx 0.732\, c(t_0)$
which is in contradiction the assumption unless $f(0)=0$. Hence for $p=1$ 
the wall imposes again a Dirichlet boundary condition. 

\subsection{$p=0, v=-\frac12$}
%
For $p<0$ the wall retracts by one unit after every second time step.
Let us first consider the continuous temporal evolution between two moves when the wall is
fixed at height ${n_0}=\lfloor -t_0/2 \rfloor$. As usual the probabilities
$P_n(t)$ with $n \leq n_0$ vanish and the he first two levels, 
where the height probability distribution is nonzero, evolve according
to the differential equations
\begin{equation}
\label{UpperDGL}
\begin{split}
&\frac{\partial}{\partial t} P_{n_0+1}(t) \;=\; \frac{P_{n_0+2}^2}{P_{n_0+2}+P_{n_0+3}} \\
&\frac{\partial}{\partial t} P_{n_0+2}(t) \;=\; -\frac{P_{n_0+2}^2}{P_{n_0+2}+P_{n_0+3}}
+\frac{P_{n_0+3}^2} {P_{n_0+3}+P_{n_0+4}} 
\end{split}
\end{equation}
Just after retraction level $n_0+1$ can be visited by the interface for the
first time, hence $P_{n_0+1}(t_0)$ is initially zero and becomes nonzero as time evolves.
Assuming that $f(0) \neq 0$ and $|f'(0)| < \infty$, to leading order
the other probabilities have the
initial values $P_{n_0+2}(t_0)  \simeq P_{n_0+3}(t_0)\simeq t_0^{-1/3} f(0) =: c\,(t_0)$.
Iterating the differential equations~(\ref{LowerDGL}) over two
time steps one obtains $P_{n_0+1}(t_0+2)\simeq c\,(t_0)$ while the higher levels remain unchanged.
Thus the almost constant probability distribution in the vicinity the wall is simply
'extended' to the new level that becomes available by retraction of the wall,
meaning that a nonzero value $f(0) \neq 0$ and $|f'(0)| < \infty$ is consistent
with the equations~(\ref{UpperDGL}). Loosely speaking the wall retracts so quickly
that it does not impose a specific boundary condition, allowing $f(0)$ to take
any positive value. The equations are built in such a way that this value is
simply copied from the following height level.



\begin{thebibliography}{99}

\bibitem{Dietrich}
S. Dietrich, in {\it Phase Transition and Critical Phenomena}, edited by
C. Domb and J.L Lebowitz (Academic Press, London, Orlando, 1988), Vol 12, p. 1.

\bibitem{KPZ}
M. Kardar, G. Parisi and Y--C Zhang, Phys. Rev. Lett.
{\bf 56} 889 (1986).

\bibitem{FVScaling}
F. Family and T. Vicsek, J. Phys. A {\bf 18}, L75 (1985).

\bibitem{EW}
S.F. Edwards and D.R. Wilkinson, Proc. R. Soc. London A {\bf 381}, 17 (1982).

\bibitem{Barabasi}
A.-L. Barabasi and H.E. Stanley, {\it Fractal Concepts in Surface Growth}
(Cambridge University Press, 1995).

\bibitem{Zhang}
T. Halpin-Healy and Y.-C. Zhang, Phys Rep. {\bf 254}, 215 (1995).

\bibitem{Zhang2}
Y.-C. Zhang, Europhys. Lett. {\bf 9}, 113 (1989).

\bibitem{Prahofer2000}
M. Pr{\"a}hofer and H. Spohn, 
Phys. Rev. Lett. {\bf 84}, 4882-4885 (2000); 
Physica A {\bf 279}, 342 (2000);
J. Stat. Phys. {\bf 108} 1071 (2002);
J. Stat. Phys. {\bf 115} 255 (2004).

\bibitem{PNG}
E. Ben-Naim, A.R. Bishop, I. Daruka and P.L. Krapivsky, J. Phys. A {\bf 31},
5001 (1998). 

\bibitem{KBM91a}
J.M. Kim, A.J. Bray and M.A. Moore, Phys. Rev. A {\bf 44}, 2345 (1991).

\bibitem{SSM}
M. Plischke, Z. R\`acz and D. Liu, Phys Rev B {\bf 35}, 3485, (1987).

\bibitem{Wetting}
        H. Hinrichsen, R. Livi, D. Mukamel, and A. Politi,
        Phys. Rev. Lett. {\bf 79}, 2710 (1997).

\bibitem{Livi}
L. Baroni, R. Livi and A. Torcini, Phys Rev. E {\bf 63}, 036226 (2001).

\bibitem{Pikov}
V. Ahlers and A. Pikovsky, Phys. Rev. Lett. {\bf 88}, 254101 (2002).

\bibitem{MN1}
Y. Tu, G. Grinstein and M.A Mun\~oz, Phys. Rev Lett. {\bf 78}, 274 (1997).

\bibitem{MN2}
M.A Mun\~oz, F. de los Santos and A. Achahbar, Brazilian J. of Physics {\bf
  33}, 443 (2003)

\bibitem{Droz}
M. Droz and A. Lipowski, Phys. Rev. E {\bf 67}, 056204 (2003).

\bibitem{MNRG}
G. Grinstein, M.A Mun\~oz, and Y. Tu, Phys. Rev Lett. {\bf 76}, 4376 (1996).

\bibitem{ASEP}
T. M. Ligget, {\it Interacting Particle Systems}, (Springer-Verlag, New York 1985).

\bibitem{Gwa92}
L.-H. Gwa and H. Spohn, Phys. Rev. Lett. {\bf 68}, 725 (1992).

\bibitem{Johansson}
K. Johansson, Comm. Math. Phys {\bf 209}, 437 (2000).

\bibitem{Abramowitz}
M. Abramowitz and I. Stegun, {\it Handbook of Mathematical Functions} 10th
printing (New York: Dover, 1972).

\bibitem{KrMHH92}
J. Krug, P. Meakin and T. Halpin-Healy, Phys. Rev. A {\bf 45}, 638 (1992).

\bibitem{Ginelli03}
        F. Ginelli, V. Ahlers, R. Livi, D. Mukamel, A. Pikovsky, 
        A. Politi, and A. Torcini,
        Phys. Rev. E {\bf 68}, 065102(R) (2003).


\end{thebibliography}
\end{document}